\documentclass[%
 reprint, 
 amsmath,amssymb,
 aps,
citeautoscript,
prb,
floatfix]{revtex4-1}
\usepackage[colorlinks=false]{hyperref}
\usepackage{float}
\usepackage{graphicx}%
\usepackage{dcolumn}%
\usepackage{bm}%
\usepackage{xcolor}
\usepackage[normalem]{ulem}
\usepackage{color,soul}
\usepackage{float} 
\begin{document}

\preprint{APS/123-QED}

\title{Integrated Quantum Optical Phase Sensor}

\author{Hubert~S.~Stokowski\textsuperscript{1}}
\author{Timothy~P.~McKenna\textsuperscript{2}}
\author{Taewon~Park\textsuperscript{1,3}}
\author{Alexander~Y.~Hwang\textsuperscript{1}}
\author{Devin~J.~Dean\textsuperscript{1}}
\author{Oguz~Tolga~Celik\textsuperscript{1,3}}
\author{Vahid~Ansari\textsuperscript{1}}
\author{Martin~M.~Fejer\textsuperscript{1}}
\author{Amir~H.~Safavi-Naeini\textsuperscript{1,4}}

\affiliation{%
\textsuperscript{1} \mbox{Department of Applied Physics and Ginzton Laboratory, Stanford University, Stanford, California 94305, USA}
}%
\affiliation{%
\textsuperscript{2} \mbox{Physics \& Informatics Laboratories, NTT Research, Inc., Sunnyvale, California 94085, USA}
}%
\affiliation{%
\textsuperscript{3} \mbox{Department of Electrical Engineering, Stanford University, Stanford, California 94305, USA}
}%
\affiliation{%
\textsuperscript{4} safavi@stanford.edu
}%

\date{\today}

\pacs{Valid PACS appear here}
\maketitle

{\bf
The quantum noise of light fundamentally limits optical phase sensors. A semiclassical picture attributes this noise to the random arrival time of photons from a coherent light source such as a laser. An engineered source of squeezed states suppresses this noise and allows sensitivity beyond the standard quantum limit (SQL) for phase detection. Advanced gravitational wave detectors like LIGO have already incorporated such sources, and nascent efforts in realizing quantum biological measurements have provided glimpses into new capabilities emerging in quantum measurement. We need ways to engineer and use quantum light within deployable quantum sensors that operate outside the confines of a lab environment. Here we present a photonic integrated circuit fabricated in thin-film lithium niobate that provides a path to meet these requirements. We use the second-order nonlinearity to produce a squeezed state at the same frequency as the pump light and realize circuit control and sensing with electro-optics. Using a 26.2 milliwatts of optical power, we measure (2.7 $\pm$ 0.2 )$\%$ squeezing and apply it to increase the signal-to-noise ratio of phase measurement. We anticipate that on-chip photonic systems like this, which operate with low power and integrate all of the needed functionality on a single die, will open new opportunities for quantum optical sensing.
}

\begin{figure*}[t]
  \begin{center}
      \includegraphics[width=\textwidth]{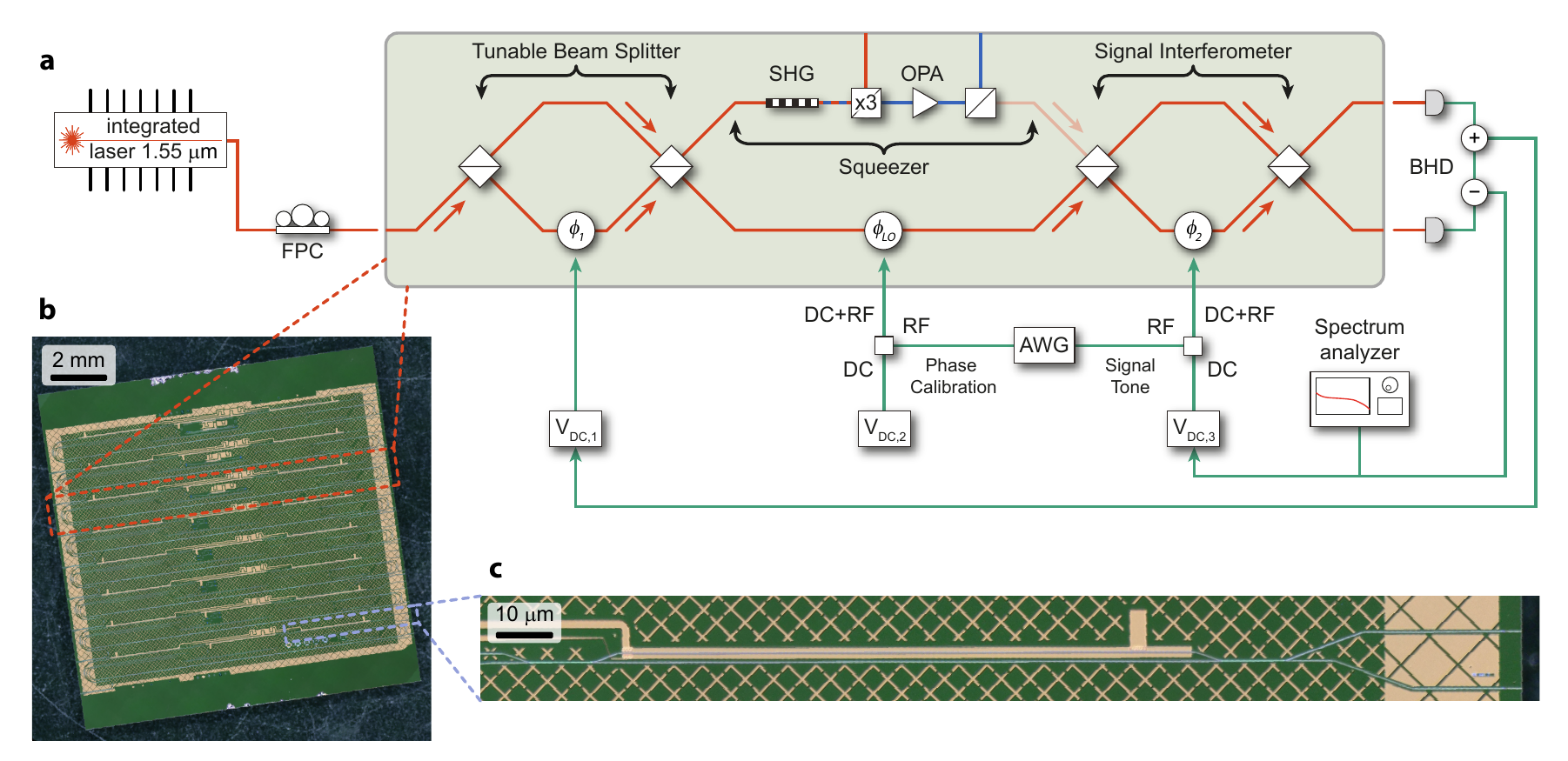}
  \end{center}
 \caption{\textbf{Squeezing and quantum-enhanced measurement PIC. }
\textbf{a},~Schematic of a photonic integrated circuit for squeezing and quantum-enhanced measurement. Light from an external, integrated laser is polarization-controlled and coupled through a cleaved facet. The PIC splits light into two paths at the tunable beam splitter: pump the squeezer and serve as the local oscillator. The two beams are recombined at the signal interferometer to perform a quantum-enhanced measurement of a small RF signal. The signal interferometer also serves as a 50/50 beam splitter for squeezing characterization in BHD. Light is detected off-chip with a balanced photoreceiver connected to the chip with two lensed multimode fibers. Sum and difference signals provide feedback for locking output splitting ratio and LO power level.
\textbf{b},~Microscope picture of a chip with eight PICs, the red rectangle highlights a single device, blue outlines a single signal interferometer. A large-scale gold cross-pattern covers the entire chip to prevent stray light from interfering with guided modes.
\textbf{c},~Microscope picture of an MZI serving as a signal interferometer. The asymmetric design provides phase modulation to one of the arms of the interferometer through gold electrodes routed to an RF probe.
FPC – fiber polarization controller, SHG – second harmonic generation, OPA – optical parametric amplification, BHD – balanced homodyne detection, AWG – arbitrary waveform generator.}
 \label{fig:fig1}
\end{figure*}

Squeezed states of light exhibit fluctuations in their quadrature amplitude below that of the quantum vacuum.~\cite{Andersen2016} This property makes them indispensable resources for increasing the sensitivity of optical measurements. In interferometry with classical light, signal-to-noise ratio scales as $\sqrt N$ with the total number of detected photons~$N$. As a result, the imprecision in determining the optical phase, or phase sensitivity, follows $1/\sqrt N$, referred to as the standard quantum limit (SQL). To increase the SNR within the confines of the SQL, we must increase $N$ -- by using more optical power or extending the measurement time. These approaches are often undesirable or impossible due to technical or fundamental limitations. Intriguingly, the SQL can be overcome by injecting squeezed states of light into the dark port of the interferometer~\cite{caves1980measurement}. Ultimately, this approach may lead to a truly quantum-limited Heisenberg scaling $1/N$ of sensitivity~\cite{caves1980measurement, giovannetti2004quantum, Giovannetti2006}. Over the past forty years, impressive efforts have been put into designing squeezed light sources and integrating them into the gravitational wave detectors. These detectors now routinely obtain quantum-enhanced sensitivity exceeding 2~dB~\cite{Goda2008_LIGO, aasi2013enhanced}. Similar ideas were also extended into other laboratory demonstrations, e.g., in distributed phase sensing~\cite{guo2020distributed} and enhanced signal-to-noise-ratio in biological sensing~\cite{Taylor2013} and microscopy~\cite{casacio2021quantum}.

We aim to demonstrate an \emph{integrated} quantum optical sensor that generates and uses squeezed light to perform measurements beyond the SQL. Our sensor moves much of the functionality often implemented on the optical table -- e.g., generating pump light, locking phases, and implementing an interferometer, onto the chip itself. We use thin-film lithium niobate (LN)~\cite{Zhang2017} to achieve all of the functions in a monolithic circuit. In the last few years, integrated thin-film LN devices have demonstrated efficient light generation covering wavelengths from mid-infrared to blue~\cite{Wang2018, Lu2020, Mishra2021, Park2022}, optical parametric oscillators~\cite{Lu2021, mckenna2022ultra}, dispersion engineering~\cite{Jankowski2020, Jankowski2021, Mishra2022}, EO modulation, tuning~\cite{Wang2018a, Celik2022}, and frequency combs~\cite{Zhang2019}. This article combines nearly all of the aforementioned capabilities in a single  LN chip and demonstrates an integrated quantum sensor. We leverage the strongest electro-optic coefficient of LN ($r_{33} = 31~\text{pm/V}$)~\cite{Boyd2008} to actively control the power and phase of light within a photonic integrated circuit (PIC) and interface it with periodically poled LN (PPLN) waveguides that support efficient nonlinear processes. The strong second-order optical nonlinear coefficient $\chi^{(2)}$, accompanied by quasi-phase-matching in PPLN, allows us to achieve efficient second harmonic generation (SHG) and squeezed light generation through optical parametric amplification (OPA).

In the degenerate parametric amplification process, photons are generated in pairs with a fixed phase relationship through the interaction $\hat{H}_{\text{I}} \propto ( \hat{a}^2 + \hat{a}^{\dagger 2})$. The resulting state of the light is squeezed. Following the first demonstration of a squeezed light source by parametric down-conversion (PDC) in an optical cavity~\cite{Wu1986}, numerous other approaches, including single-pass PDC~\cite{kaiser2016fully, kashiwazaki2020continuous}, semiconductor lasers, optical fibers, atomic ensembles~\cite{Andersen2016}, and ponderomotive squeezing with mechanical oscillators~\cite{SafaviNaeini2013} have successfully demonstrated varied amounts of squeezed light. Optical materials with intrinsic second- or third-order nonlinearities may be the most versatile integrated sensors approach. Recent demonstrations of integrated photonic circuits in silicon nitride~\cite{Vaidya_2020, Zhang2021} and thin-film LN~\cite{Dutt2015, Chen2022, Nehra2022} have shown that low power and scalable sources of quantum light are possible.

From a photonic system perspective, the squeezed source is only one part of a larger optical sensor. Combining emerging squeezed light sources with more complex PICs is a promising avenue for developing deployable optical sensors with quantum-enhanced sensitivity. So far, achieving sub-shot-noise sensitivity in optical metrology has required complex setups with multiple modulators, lasers, and optical cavities. This has limited the domain of quantum optical sensing to experiments with high complexity~\cite{Goda2008_LIGO, aasi2013enhanced, casacio2021quantum} that can address only a few of the possible application spaces -- even for experiments that use new integrated quantum light sources. Our work combines emerging squeezed light sources with a complex PIC to realize a nearly complete on-chip quantum sensor, thus opening a promising avenue for deployable optical sensors with quantum-enhanced sensitivity.

\section{Results}

In this work, we use the X-cut thin-film of lithium niobate on insulator (LNOI) platform to build a sensor with an integrated source of squeezed light for quantum-enhanced measurement. Fig.~\ref{fig:fig1}a outlines a scheme of the PIC implemented in our chip with three major components – input tunable beam splitter, squeezer, and signal interferometer. We pattern eight such PICs in a single chip, which we show on a microscope picture in Fig.~\ref{fig:fig1}b. We design the squeezer to generate a sub-shot noise state of light at the same frequency as the pump light, enabling interference with the local oscillator (LO) extracted from the original beam. Active EO circuitry controls interferometers and the LO phase, which enables using the system for both squeezing characterization and quantum measurement. Using lensed SMF-28 fiber, we interface our device with an off-the-shelf DBR laser and operate the system in CW mode at 1544 nm. The output ports of the beamsplitter out-couple light into two lensed multimode fibers, which we connect to a balanced photoreceiver off-chip. Both the laser~\cite{Beeck2021, Li2022} and detectors~\cite{tasker2021silicon} can be further integrated with the PIC to increase the system efficiency and improve the squeezing visibility.

\subsection{Device Design}

The squeezer subsystem comprises two PPLN waveguides and a series of directional couplers for spectral filtering. FH light first enters a PPLN waveguide designed to phasematch a second harmonic generation process using periodic poling with a period of around 3.7~\textmu m. Tight mode confinement and strong second-order optical nonlinear coefficient of LNOI yield highly efficient SHG. Next, we filter out the the residual FH by passing the light through three directional couplers. By design, each directional coupler transfers majority of FH while keeping the second harmonic (SH) light in the original waveguide; hence the structure acts as a dichroic beamsplitter. The SH light then loops back to another PPLN waveguide, realizing a phase-sensitive OPA at the FH frequency, squeezing one quadrature and anti-squeezing the other. The amount of squeezing generated in this structure is given by:
\begin{eqnarray}
\cfrac{\langle\delta \hat{\mathrm{X}}^2\rangle} {\langle\delta \hat{\mathrm{X}}_{\text{vac}}^2\rangle}
=
\text{exp} \Bigg(
{-2 L \, \sqrt{\eta P_{\text{in}}}\, \text{tanh} \Big( L\, \sqrt{\eta\, P_{\text{in}}} \Big) } 
\Bigg),
\label{eq:eq1}
\end{eqnarray} where $\langle \delta \hat{\mathrm{X}}^2 \rangle$ is the variance of the quadrature squeezed with respect to vacuum $\langle \delta \hat{\mathrm{X}}_{\text{vac}}^2 \rangle$, $L$ is the length of the poled waveguides, $\eta$ is the normalized nonlinear conversion efficiency~\cite{Wang2018}, and $P_{\text{in}}$ is the telecom pump power (see Methods for details). Finally, the SH light is filtered out using the same directional coupler design, and the squeezed state propagates to the signal interferometer at the output of the PIC. We route the light filtered out after the SHG and OPA sections to the chip facet and use them as monitor ports.

Active EO circuitry controls optical power splitting and the LO phase within the PIC. The \emph{tunable beam splitter} at the input and the \emph{signal interferometer} at the output are both intensity modulators. Both consist of a Mach-Zehnder interferometer (MZI) with a phase modulator in one arm. The former is typically DC-biased at phase $\phi_1$ either close to zero, to direct most of the input light into the squeezer, or set to $\pi$, to send most of the light into the LO and characterize shot noise. The latter MZI, shown in Fig \ref{fig:fig1}c, is always locked with a DC bias to make the power at the two outputs equal for a balanced homodyne detection (BHD) measurement. We use the same phase modulator design for quadrature selection in the squeezer path through $\phi_{\text{LO}}$ control. In addition to DC bias, we apply RF tones to the LO phase modulator and the signal interferometer. The RF phase modulation on the LO serves as voltage-to-phase calibration and quantifies the light leakage in the squeezer $\sqrt{\varepsilon}~=~\alpha_{\text{LO}}/\alpha_\text{Leakage}$, while the RF phase modulation imposed on the signal interferometer forms the signal we measure.

\begin{figure}
  \begin{center}
      \includegraphics[width=1\columnwidth]{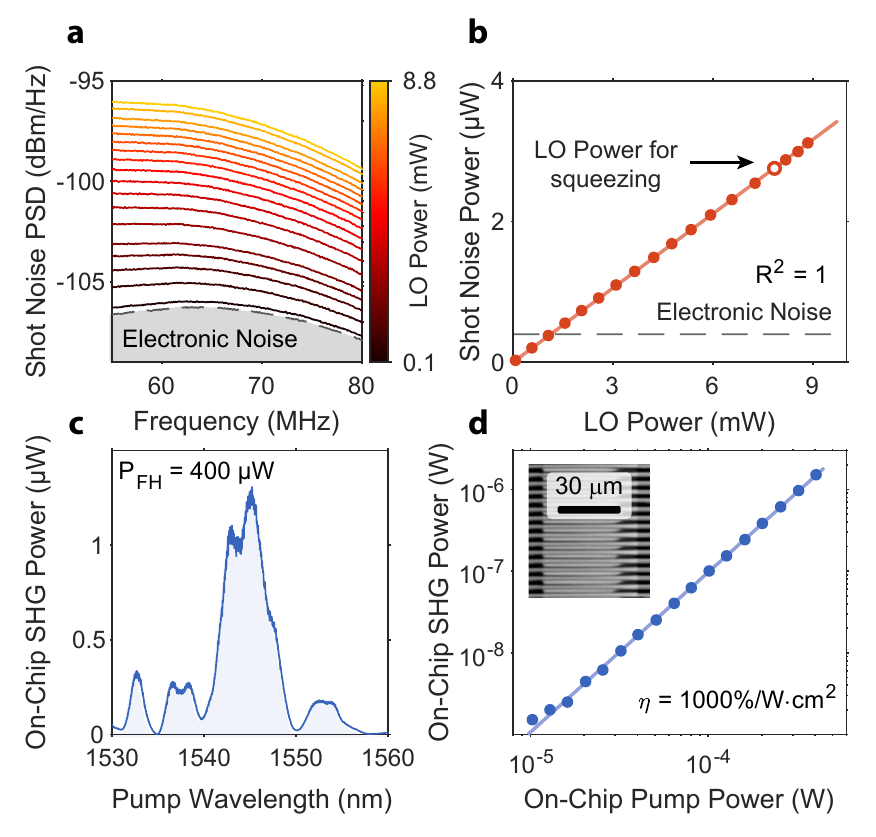}
  \end{center}
 \caption{
 \textbf{PIC subsystems characterization. }
\textbf{a},~Shot noise power spectral density for different LO powers on the chip. We characterize our setup to find LO power yielding signals about 10 dB above the electronic noise floor.
\textbf{b},~Integrated noise from Fig.~\ref{fig:fig2}a vs. LO power follows a perfect linear trend, proving that the system is limited by shot noise at the target LO power of 7.8 mW.
\textbf{c},~Example SHG phasematching curve, deviation from ideal $\text{sinc}^2$ shape results from small geometry variation along the waveguide length.
\textbf{d},~Measured on-chip SHG peak power vs. FH pump power. Linear fit yields normalized efficiency of the nonlinear process of about $1,000 \%/(\mathrm{W\cdot cm}^2)$. The inset shows SHG microscope image of high-quality periodic poling of the thin-film lithium niobate. We pattern waveguides in the area where poling extends through the entire thickness of the LN film.}
 \label{fig:fig2}
\end{figure}

\begin{figure*}
  \begin{center}
      \includegraphics[width=\textwidth]{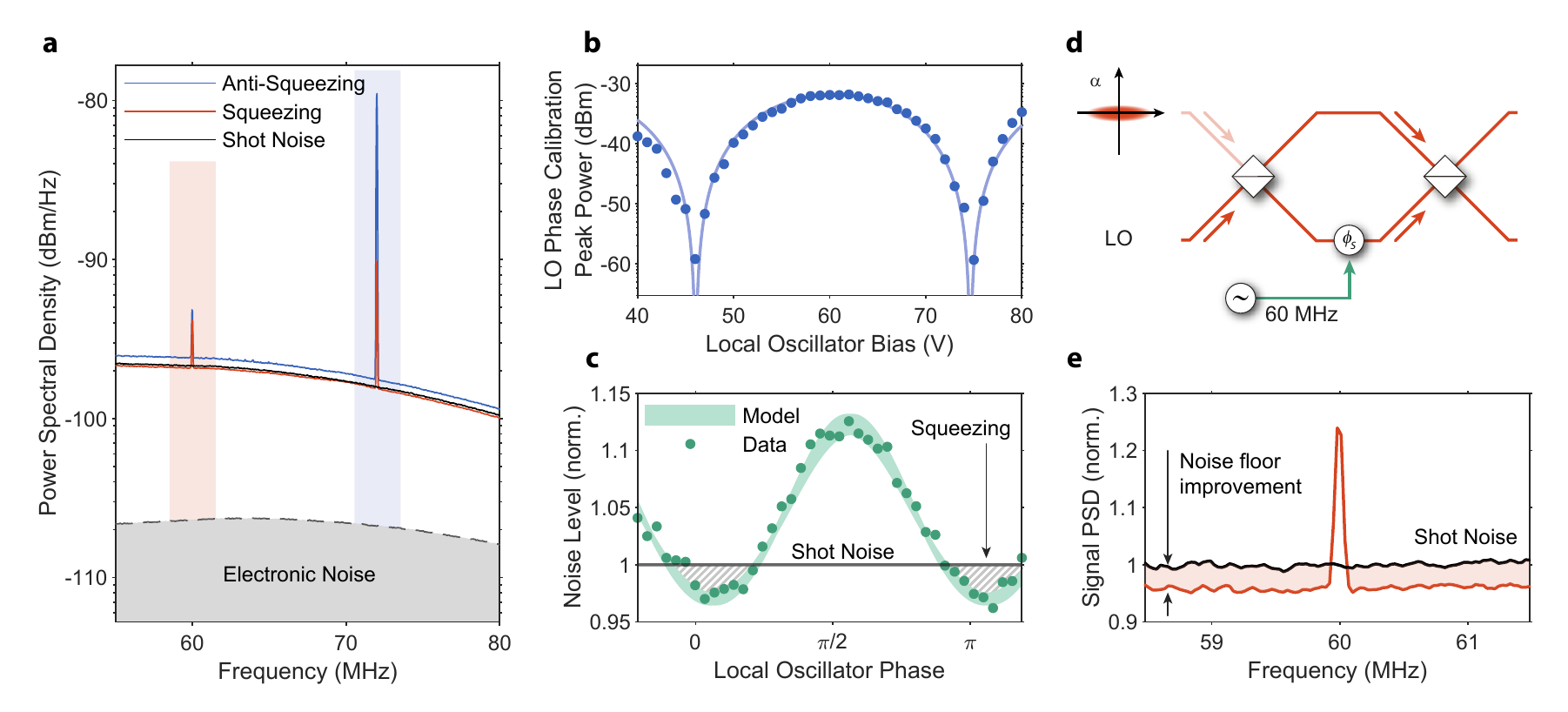}
  \end{center}
 \caption{
 \textbf{Squeezing and quantum-enhanced detection. }
\textbf{a},~Power spectral density for maximum squeezing (orange line) and maximum anti-squeezing (blue line), collected while pumping the squeezer with 18.4 mW of FH. The black line provides a shot noise measurement reference. We highlight two RF peaks resulting from modulation of the LO phase shifter (blue highlighted area) and signal interferometer (orange highlighted area).
\textbf{b},~Local oscillator calibration peak power vs. DC bias voltage of the phase shifter. The solid line indicates the fit line, from which we find $V_{\pi} = 28.6 \text{ V}$ and light leakage into squeezing path $\varepsilon \approx 4.0\%$.
\textbf{c},~Integrated noise power as a function of the LO phase (calibrated in Fig.~\ref{fig:fig3}b). The grey-dashed area represents the characteristic sub-shot noise behavior of a squeezed state. Misalignment of the squeezed state and LO phases is caused by the path length difference between the two paths. The green highlighted area represents the model, including squeezer leakage, detection efficiency, and normalized efficiency. The uncertainty of the normalized efficiency and phase sets the model bounds.
\textbf{d},~Quantum-enhanced measurement scheme – squeezed state feeds into a dark port of the MZI along with the LO entering the bright port. Phase modulation applied in one of the arms can be detected with enhanced SNR compared to a classical measurement.
\textbf{e},~Measurement of a small RF tone with noise floor suppressed with squeezing. We align the LO phase with maximum squeezing by DC tuning and measure the RF power spectrum through a BHD. For a $-83$~dBm modulation, we observe about (4 $\pm$ 1)$\%$ improvement of the SNR compared to a shot-noise limited measurement.}
\label{fig:fig3}
\end{figure*}
\subsection{Subsystem Characterization}

We calibrate the shot noise level by routing all the optical power to the local oscillator path. For this purpose, we bias the tunable beam splitter at the input with a DC voltage bias of 47~V. We achieve this by minimizing the amount of light in the squeezer's fundamental harmonic (FH) monitor port. In this configuration, we use the local oscillator to probe the vacuum state sent to the opposing port of the signal interferometer. For the BHD measurement, we lock the signal interferometer to a splitting ratio of 50:50. We achieve this by controlling the difference between the intensity of two outputs and locking its DC part to zero. Next, we step the power of the input laser using an external variable optical attenuator (VOA), such that the power in the local oscillator waveguide varies between $0$ and $8.8~\text{mW}$ and record the RF spectra of the balanced homodyne detector, as shown in Fig.~\ref{fig:fig2}a. Finally, we integrate the noise power and plot it against the LO power in Fig.~\ref{fig:fig2}b, observing a linear dependence as expected from a shot noise-limited measurement. The mean squared error of the linear fit defines our uncertainty of the shot noise measurement. Based on the shot noise calibration, we choose LO power for the squeezing experiments at around $7.8~\text{mW}$, where the optical shot noise dominates our detection. We highlight this operating point in Fig.~\ref{fig:fig2}b.

The squeezer relies on the performance of the periodically poled \mbox{waveguides} in the SHG and OPA sections. We characterize our periodic poling by measuring the quasi-phase-matching transfer function of a diagnostic \mbox{waveguide} patterned adjacent to our SHG and OPA segments such that it passes through the same poling region (see Methods for measurement details). An example QPM transfer function is shown in Fig.~\ref{fig:fig2}c. This measurement gives us the SHG response decoupled from the rest of the PIC components. Note that the diagnostic waveguide is 8.1 mm long, whereas the one in the primary device is 10 mm long. Given the significant length of the fabricated waveguide, nonuniformities distort the ideal $\mathrm{sinc}^2$ shape of the nonlinear transfer function~\cite{Kuo2021} and reduce the peak nonlinear efficiency. We measure the intensity of generated SH light as a function of pump power controlled by external VOA. In Fig.~\ref{fig:fig2}d, we plot data and linear fit to extract a maximum normalized efficiency of about $1,000~ \%/(\mathrm{W\cdot cm}^2)$, lower than the simulated $4,000~ \%/(\mathrm{W\cdot cm}^2)$ for an ideal poled structure. We use this number further to estimate the amount of parametric amplification and the expected squeezing level. Because of the nonuniformities, the quadratic scaling of the efficiency with length is not exact. However, we neglect this because the length difference between the test and device waveguides is small. The inset of Fig.~\ref{fig:fig2}d shows a second harmonic microscope image of periodic poling used in the fabrication process. Black stripes on the sides of the picture are metal finger electrodes, and light grey shapes stretching between them correspond to inverted crystal domains.

\subsection{Squeezing and Quantum-Enhanced Sensing}

We characterize squeezing by directing most of the laser light into the squeezer and holding the LO power constant at around $7.8~\text{mW}$. In this case, we drive the device with $125~\text{mW}$ of optical power off-chip, of which $18.4~\text{mW}$ drives the SH generation directly. Given the calibrated normalized efficiency, we estimate SH power at around $3.4~\text{mW}$ to generate about (1.5 $\pm$ 0.1) dB of squeezing on the chip. We perform a BHD measurement to verify this by beating the squeezed state with the local oscillator at the signal interferometer. We lock the phase shift within the output interferometer such that the DC part of the difference signal is zero. At the same time, we lock the input beam splitter to keep the sum of photodiode readouts at a constant value. We follow this locking scheme to work around the slow phase drift observed in EO devices~\cite{Zhu2021, Wang2022}. A simple procedure that periodically updates the voltage at the input beam splitter and signal interferometer at a low frequency (around 1~Hz) is sufficient. Finally, we measure the power spectral density, as shown in Fig.~\ref{fig:fig3}a, and reference it to the shot noise measurement (black line). We perform this measurement as a function of the LO phase, controlled through a DC voltage applied to a phase shifter in the LO path. Blue and orange traces in Fig.~\ref{fig:fig3}a correspond to the maximum anti-squeezing and maximum squeezing, respectively. In addition to probing broadband noise, we introduce two RF signals corresponding to peaks at 72~MHz and 60~MHz. The former originates from an RF signal applied to the LO phase shifter, which we highlight in blue, and the latter corresponds to a tone on the signal interferometer, which we highlight in orange.

We verify the phase-sensitive nature of the generated squeezed light by stepping the LO phase shifter bias. Due to a small power leakage of the FH in the squeezer path, $\varepsilon$, we can observe the RF peak introduced at the LO phase shifter at 72 MHz. This signal is proportional to $\varepsilon \, \text{sin}^2(V_{DC}/V_{\pi} \cdot \pi)$ (see Methods for details). By fitting the peak power as a function of LO bias voltage in Fig.~\ref{fig:fig3}b, we establish the power leakage ratio $\varepsilon$ at around (4.0~$\pm$~1.0)$\%$, which we use to estimate the filter extinction ratio at around 18 dB. This amount of leakage into the squeezer enables our LO phase calibration procedure, but negatively impacts the squeezing visibility. Using the fit, we also find the half-wave voltage of our LO to be 28.6~$\pm$~0.3~V and convert the DC bias to phase. The half-wave voltage agrees with our design (see Methods for details). In the squeezing measurement, in Fig.~\ref{fig:fig3}c, we plot the integrated noise floor over the measurement bandwidth vs. calibrated LO phase, excluding the two RF peaks and a small, broad feature between 67 and 72 MHz. Plotted noise is normalized to shot noise, in which uncertainty is smaller than the width of the line. We see expected periodic behavior, where the minimum and maximum correspond to probing squeezed and anti-squeezed field quadratures, respectively. Instead of performing fast phase sweeps, we maintain the LO phase at each tuning point for roughly 13 seconds and obtain the full RF spectrum. This highlights the stability of our system, a necessary ingredient for a stable and metrologically useful source of squeezed light. We measure (2.7~$\pm$~0.2)~$\%$ of squeezing and (12.3~$\pm$~0.2)~$\%$ of anti-squeezing. Limited squeezing visibility and squeezing/anti-squeezing asymmetry result from the detection chain efficiency of 20$\%$ (see Methods for more details) and squeezer leakage $\varepsilon \approx 4\%$. Accounting for these factors, we expect the level of squeezing on-chip to be around (1.5~$\pm$~0.1)~dB, which agrees with our estimation based on the normalized efficiency. Total detection efficiency can improve up to 83$\%$ by reducing optical loss. Implementing straightforward PIC layout improvements to reduce the propagation loss, improved fiber-to-chip coupling~\cite{He2019, Hu2021, Ying2021}, and higher quantum efficiency photodiodes can result in an achievable 4.5 dB squeezing or SNR improvement by 280$\%$ using the same off-the-shelf integrated laser. See the Methods section for more details.

Finally, we use the signal interferometer to perform a quantum-enhanced measurement of an RF signal. In this case, the chip configuration is the same as in the squeezing measurement; most light pumps the squeezer, and the LO power is fixed at $7.8~\text{mW}$. We use the signal interferometer to perform a quantum-enhanced measurement by injecting squeezed state into its dark port, as shown in Fig.~\ref{fig:fig3}d. Based on our squeezing calibration, we set the LO phase to align with the maximum observed squeezing and perform the measurement. For that purpose, we apply a small RF tone with an amplitude of -83~dBm, corresponding to the root-mean-square voltage of 15.8~\textmu V at 60 MHz. This signal can be detected in the BHD measurement, as shown in Fig.~\ref{fig:fig3}e, and it is not sensitive to the LO phase. As a result, the measurement noise floor is reduced with respect to the shot noise and results in about \mbox{(4~$\pm$~1)$\%$} SNR improvement. This data was taken at a different time than the LO phase sweep dataset shown in Fig~\ref{fig:fig3}c, and so we attribute the slightly higher level of observed squeezing to slightly different experimental conditions.

\section{Discussion}

Complex quantum photonic integrated circuits, like the one presented in this work, extend the reach of quantum technologies into new domains. The integration allows us to generate quantum light and leverage its properties within the same circuit, enabling more stable and lower power operation. These critical advantages point toward new classes of mobile and deployable sensors that are natively ``quantum-compatible''. In the work we present here, we benchmark a sensor by detecting voltage signals imparted as phase upon an optical field. Many other interferometric measurements are today limited by shot noise. These measurements, including refractive index sensing, optomechanical force, acceleration~\cite{Krause2012}, and mass sensing~\cite{Hanay2012}, will directly benefit from integrating the quantum apparatus, as demonstrated here. Moreover, squeezing and squeezed light are crucial elements of numerous quantum sensing, communications\cite{Gehring2015}, and computing protocols~\cite{Zhong_2020, Zhong2021}. Quantum PICs, such as those shown here, are promising and scalable approaches for building these systems.

\section{Data Availability}
The data sets generated during and/or analyzed during this study are available from the corresponding author on request.

\section*{Author contributions}
H.S.S. designed the device. 
H.S.S., T.P., and A.Y.H. fabricated the device.
H.S.S., V.A., T.P.M., and O.T.C. developed the fabrication process.
M.M.F. and A.H.S.-N. provided experimental and theoretical support.
H.S.S., V.A., T.P., and D.J.D. performed the experiments and analyzed the data.
H.S.S., V.A., and A.H.S.-N. wrote the manuscript.  
H.S.S., V.A., T.P.M, and A.H.S.-N. conceived the experiment, and A.H.S.-N. supervised all efforts. 

\section*{Competing interests}
The authors declare no competing interests.

\section*{Acknowledgements} H.S. and V.A. thank Kevin Multani, Christopher Sarabalis, and Michael Stefszky for discussions and technical support. The authors wish to thank NTT Research for financial and technical support. This work was supported by the DARPA Young Faculty Award (YFA). Part of this work was performed at the Stanford Nano Shared Facilities (SNSF) and Stanford Nanofabrication Facility (SNF), supported by the National Science Foundation under award ECCS-2026822.  H.S. acknowledges support from the Urbanek Family Fellowship. V.A. acknowledges support from Stanford Q-FARM Bloch Fellowship. D.D. acknowledges support from the NSF GRFP (No. DGE-1656518).

\bibliography{2022_squeezer_bibliography}

\newpage

\section*{Methods}

\section*{Fabrication} 
\begin{figure*}
  \begin{center}
      \includegraphics[width=1\textwidth]{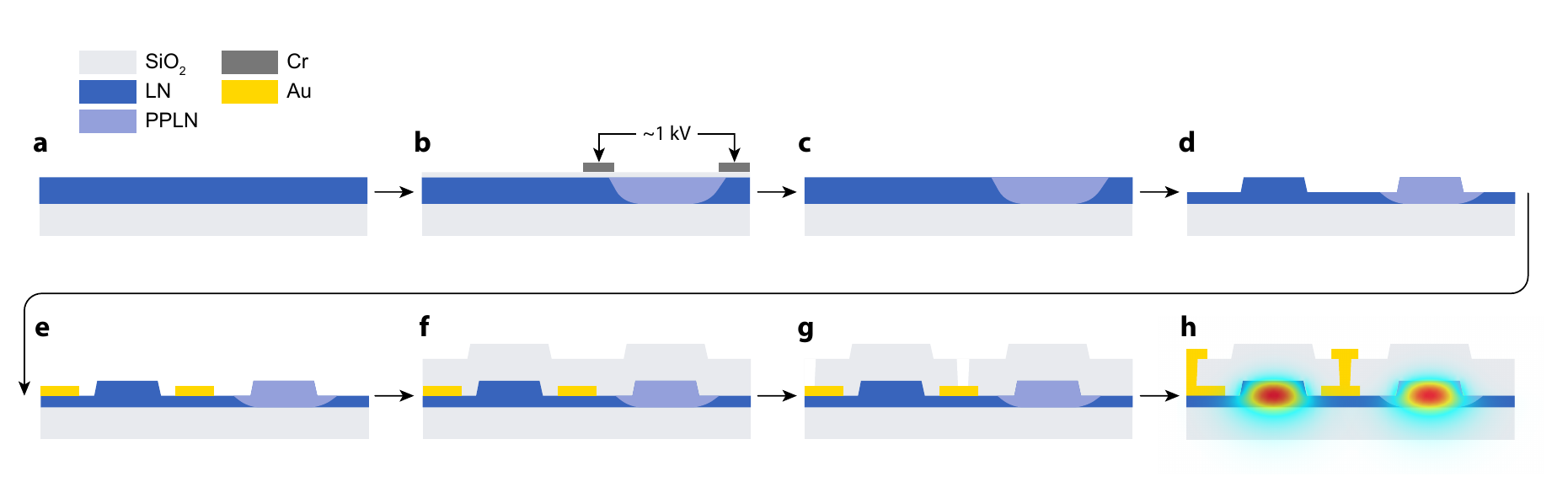}
  \end{center}
 \caption{
 \textbf{Device fabrication procedure. }
\textbf{a},~We start with a 500 nm thin-film LN on insulator.
\textbf{b},~For periodic poling of the LN film, we use chromium electrodes deposited on a 100 nm silica insulation layer. Next, we apply a high-voltage pulse to invert crystal domains.
\textbf{c},~Chromium electrodes and silica layer are removed after poling using chromium etchant and buffered oxide etchant.
\textbf{d},~Waveguides are patterned using electron-beam lithography and argon ion milling.
\textbf{e},~We pattern a bottom layer of gold directly on LN to increase the electric field within the waveguide region.
\textbf{f},~Wavegues are covered with around 700~nm of silica. 
\textbf{g},~We open vias in the cladding to provide electrical contact to the buried electrodes.
\textbf{h},~We pattern the top layer of gold to connect our devices to the external probe.
}
\label{fig:figSI1}
\end{figure*}

We fabricate our device with X-cut thin-film LNOI (lithium niobate on insulator, NanoLN), following the procedure outlined in figure \ref{fig:figSI1}. The material stack consists of 500~nm of lithium niobate bonded to a 2~\textmu m-thick silica layer on top of a silicon handle wafer, as shown in Fig.~\ref{fig:figSI1}a. Fabrication starts with poling the thin-film. In the first step (Fig.~\ref{fig:figSI1}b), we pattern Cr electrodes on an insulating layer of 100~nm SiO$_2$ using electron beam lithography (JEOL~6300-FS,~100-kV) and liftoff process. Poling period is around 3.7~\textmu m, and we design it for phasematching between our waveguide's 1550~nm and 775~nm modes. Next, we apply high voltage pulses to flip crystal domains~\cite{Nagy2019, mckenna2022ultra}. After poling, we remove the electrodes with chromium etchant and buffered oxide etchant (Fig.~\ref{fig:figSI1}b).

The photonic circuit is patterned with a FOX-16 mask and electron beam lithography and transferred to the LN layer with an argon ion mill (Fig.~\ref{fig:figSI1}d). The etch depth is 300 nm, and the waveguide width is 1.2 \textmu m.

Metal electrodes for EO tuning are patterned with the same process as the poling electrodes but made out of 100 nm gold (Fig.~\ref{fig:figSI1}e). The bottom layer of electrodes and waveguides are covered with SiO$_2$ deposited with a high-density plasma process (Fig.~\ref{fig:figSI1}f)~\cite{ShamsAnsari2022}. To provide electrical contact to the buried electrodes, we pattern vias by standard SiO$_2$ etch with fluorine chemistry and photolithography (Fig.~\ref{fig:figSI1}g). Finally, we pattern a top metal layer with the electron beam. This layer provides access to the buried electrodes with an external probe and consists of 200 nm of gold and a 10 nm chromium adhesion layer (Fig.~\ref{fig:figSI1}h).

In the final step, we prepare the chip facets for light coupling by stealth dicing with a DISCO DFL7340 laser saw~\cite{Jankowski2020}. High-energy optical pulses are focused on the substrate to create an array of damage sites. They act as nucleation sites for crack propagation, resulting in a uniform and smooth cleave.

\section*{Electro-Optic simulation} 

\begin{figure}
  \begin{center}
      \includegraphics[width=1\columnwidth]{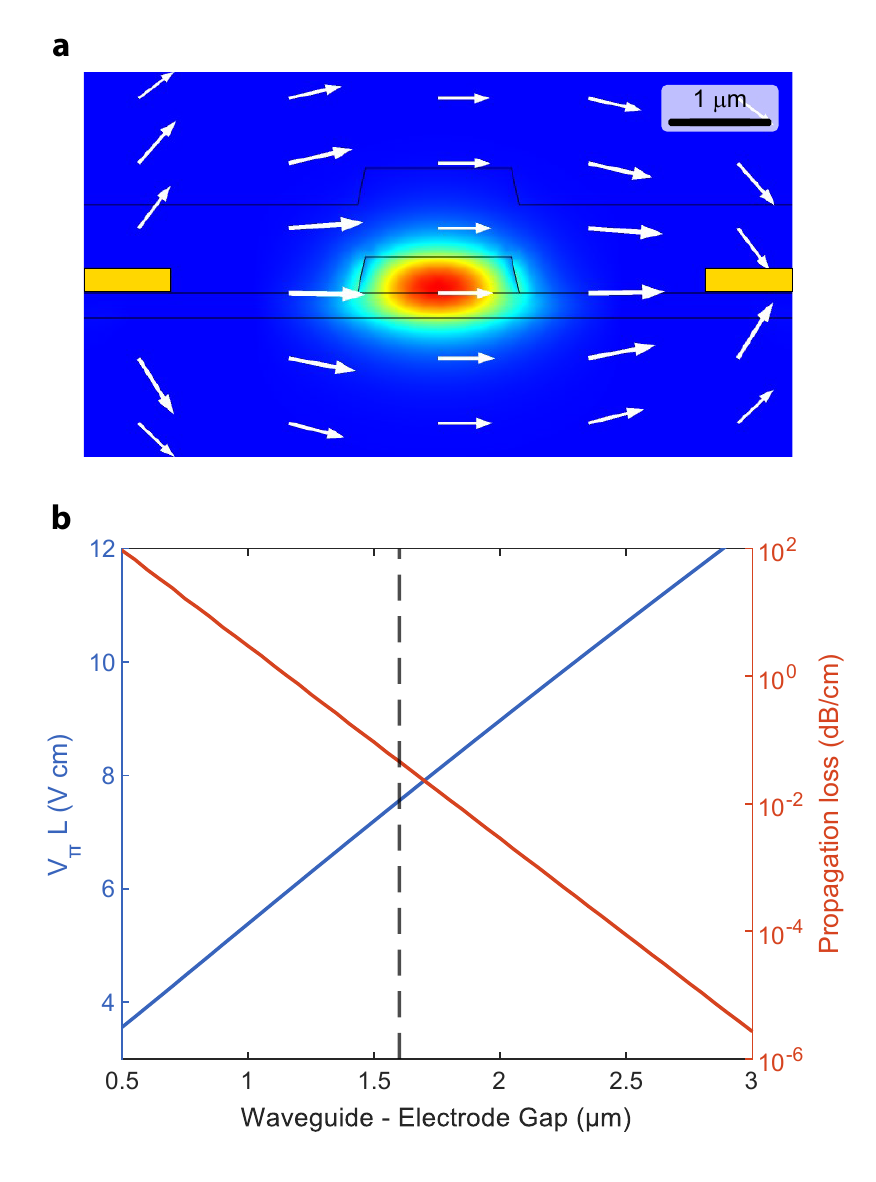}
  \end{center}
 \caption{
 \textbf{Design of the electro-optic modulator. }
\textbf{a},~Eigenmode solver solution for the fundamental TE mode in the presence of a static electric field. White arrows indicate the direction of the applied bias field.
\textbf{b},~Half-wave voltage-length product and metal-induced propagation loss as a function of the waveguide-electrode gap. The dashed line corresponds to the geometry described in the main text with a waveguide-electrode gap of 1.6 µm.
}
\label{fig:figSI2}
\end{figure}

We design the geometry of our electro-optic devices to provide phase-tuning functionality without introducing excess loss to the quantum state of light. We model our system using a finite-element mode solver (COMSOL). We first define the electric-field-dependent refractive index~\cite{Boyd2008} as:
\begin{eqnarray}
n_{\text{o}}'=n_{\text{o}}
-
\frac{1}{2}r_{13}n_{\text{o}}^3E_{\text{z}}
\\
n_{\text{e}}'=n_{\text{o}}
-
\frac{1}{2}r_{33}n_{\text{e}}^3E_{\text{z}}.
\label{eq:eo_refr_indices}
\end{eqnarray} The elements of the electro-optic tensor $r_{33} = 31~\text{pm/V}$ and $r_{13} = 10~\text{pm/V}$ modify ordinary and extraordinary indices of the crystal as a result of the static field. We solve for the static electric field and follow with eigenmode analysis in the system with modified refractive index, as shown in Fig.~\ref{fig:figSI2}a.

We find the half-wave-voltage-length product from the relationship between effective index and applied voltage $V_{\pi}\, L = \lambda/({\partial n_{\text{eff}}/\partial{V}})$. We plot the expected values along with the propagation loss from metal proximity in Fig.~\ref{fig:figSI2}b. In the device described in the main text, we use a waveguide-electrode gap of 1.6 µm and an electrode length of 2.5 mm. We expect the half-wave voltage in this geometry to be around 30 V, which agrees with the value measured in the experiment. This gap size allows us to keep the propagation loss induced by the metal below 0.05 dB/cm, which is negligible compared to the measured loss resulting from fabrication imperfections around 0.7 dB/cm.

\section*{Experimental Setup} 
We characterize fabricated PICs in two different setup configurations, one for BHD and one for SHG characterization. In both cases, the chip temperature is controlled with a thermistor (Thorlabs TH10K) and thermo-electric cooler (Thorlabs TECF2S, MTD415TE, MTDEVAL1). We operate our device at 45 $^{\circ}$C to maximize SHG intensity at the low noise laser operating wavelength.

The setup for BHD measurements is shown in Fig.~\ref{fig:fig1}a. For squeezing and quantum-enhanced measurements, we use an ultra-low noise, integrated DFB laser with an emission wavelength of around 1544 nm (Thorlabs ULN15PC). The laser is followed by a high extinction ratio fiber isolator (Thorlabs IOT-G-1550A) and a fiber polarization controller. We introduce an additional variable optical attenuator (HP 8156A) for the shot noise measurement to control the laser power. Then, we couple light to the chip with a lensed SMF-28 fiber (OZ Optics). We collect light into a balanced photoreceiver (Thorlabs PDB425C with RF output conversion gain $1.25 \cdot 10^5$ V/W) using multimode fibers lensed (in-house) and mounted on a v-groove chip with a pitch matching the PIC output waveguide separation. We collect RF spectra with an electronic spectrum analyzer (Rohde $\&$ Schwarz FSW26) with a resolution bandwidth set to 60 kHz. Tap ports for FH and SH light filtered within the squeezer subsystem are collected into another lensed multimode fiber, separated by a combination of a 50/50 splitter and optical edgepass filters, and monitored with photodiodes: Newport 1623 Nanosecond Photodetector for FH and Thorlabs APD440A for SH. Electro-optic components on the chip are controlled with DC voltage supplied with three TTi PLH250-P sources. LO phase modulator and the signal interferometer receive RF signal from an arbitrary waveform generator (Rigol DG4102). We mix RF signals and DC bias with bias-tees (ZFBT-4R2G+) and connect to the chip with a triple ground-signal-ground (GSG) probe. We estimate the total insertion loss of RF input at $1.22$~dB by combining the loss of two SMA cables (FLC-4FT-SMSM+), DC block (Inmet 8039), and the bias-tee.

For the second harmonic characterization, we reconfigure the setup to work with a continuously tunable c-band laser (Santec TSL-550, 1480–1630 nm) passing through a variable optical attenuator (HP 8156A). Next, we tap off 5$\%$ of the light for power calibration using a power meter (Newport 918D-IR-OD3R) and pass the light through a polarization controller. Light couples to the chip through an SMF-28 lensed fiber and is collected by the same kind of fiber at the output facet. Then, we outcouple light into free space and split with a 1000~nm short-pass dichroic mirror (Thorlabs DMSP1000) and filter to avoid cross-talk. Finally, we measure light intensity with avalanche photodiodes (Thorlabs APD410A and Thorlabs APD410). We use variable optical attenuators before APDs to prevent saturation and increase the dynamic range of our measurement (HP 8156A and Thorlabs FW102C).

\section*{Squeezer Performance}
The integrated squeezer proposed in the main text takes advantage of the second-order nonlinearity in LN waveguides. In this system, both SHG and OPA can be described with coupled-mode equations for the degenerate three-wave mixing process:~\cite{Boyd2008}
\begin{eqnarray}
\cfrac{\partial }{\partial z} A(z) &=& - i \sqrt{\eta}\, B(z) A^{\ast}(z)\, e^{-i\Delta k z}, \label{eq:ode1}
\\
\cfrac{\partial }{\partial z} B(z) &=& - i \sqrt{\eta}\, A^{2}(z)\, e^{i\Delta k z}. \label{eq:ode2}
\end{eqnarray} where $A(z)$ and $B(z)$ are amplitudes of the FH and SH modes with frequency relationship $2\omega_A = \omega_B$. $\eta$ corresponds to normalized efficiency, $\Delta k = 2k_A - k_B$ is the phase mismatch between FH and SH modes, which we compensate for with periodic poling. We solve these equations to obtain the magnitude of the power normalized amplitude of the generated SHG, assuming perfect phase matching:
\begin{eqnarray}
|B(z)| &=& |A(0)| \text{tanh} \Big( \sqrt{\eta}\, |A(0)| z \Big).
\label{eq:SHGampl}
\end{eqnarray}

In the squeezer subsystem, FH light is filtered out after pumping the SHG, and the SH stays in the waveguide to generate squeezing through OPA. To find the amount of squeezing generated, we will first define field quadratures for the FH mode as follows:
\begin{eqnarray}
\mathrm{X} &=& \cfrac{{A} + {A}^{\ast}}{2},
\label{eq:Xquad}
\\
\mathrm{Y} &=& i \, \cfrac{{A} - {A}^{\ast}}{2}.
\label{eq:Yquad}
\end{eqnarray} Next, we can use the equation \ref{eq:ode1} to find the evolution of the field quadratures $\mathrm{X}$ and $\mathrm{Y}$ along the waveguide length as follows:
\begin{eqnarray} \nonumber
\partial_z \Big( A(z) + A^{\ast}(z) \Big) 
&=& - i \sqrt{\eta} \Big(
B(z) A(z)^{\ast} - B(z)^{\ast} A(z)
\Big)
\label{eq:ode_X_deriv}
\\ 
&=& - \sqrt{\eta} |B(z)|\Big(
 A(z) + A(z)^{\ast}
\Big).
\end{eqnarray} In the last line we have chosen to set the phase of $B(z)$ to $-\pi/2$. Performing a similar calculation for the time evolution of the $\mathrm{Y}$ quadrature, we arrive at:
\begin{eqnarray}
\partial_z \mathrm{X} &=& - \sqrt{\eta} |B(z)| \mathrm{X},
\label{eq:ode_X}
\\ 
\partial_z \mathrm{Y} &=& \sqrt{\eta} |B(z)| \mathrm{Y}.
\label{eq:ode_Y}
\end{eqnarray} Solving these equations yields:
\begin{eqnarray}
\mathrm{X} &=& \mathrm{X}(0) \, \text{exp}({-\sqrt{\eta} |B(z)| z}),
\\
\mathrm{Y} &=& \mathrm{Y}(0) \, \text{exp}({\sqrt{\eta} |B(z)| z}).
\end{eqnarray}
It is clear that the $\mathrm{X}$ quadrature is deamplified while $\mathrm{Y}$ quadrature is amplified. This amplification and deamplification process, when applied to fields that have vacuum fluctuations, leads to anti-squeezing and squeezing of the quadrature fluctuations. In other words, the variance for quadrature operators under the action of amplification/deamplification is given by:
\begin{eqnarray}
\langle \delta \hat{\mathrm{X}}^2 \rangle &=& \text{exp} ({-2\,\sqrt{\eta}\, |B(z)|\, z})\, \langle \delta \hat{\mathrm{X}}_0^2 \rangle, \label{eq:varXexp}
\\
\langle \delta \hat{\mathrm{Y}}^2 \rangle &=& \text{exp} ( {2\,\sqrt{\eta}\, |B(z)|\, z} )\, \langle \delta \hat{\mathrm{Y}}_0^2 \rangle, \label{eq:varYexp}
\end{eqnarray} where $\langle \delta \hat{\mathrm{X}}_0^2 \rangle$ and $\langle \delta \hat{\mathrm{Y}}_0^2 \rangle$ are variances of the quantum vacuum state. In the proposed PPLN squeezer subsystem of length $L$, combining equations \ref{eq:SHGampl} and \ref{eq:varXexp} defines squeezing as:
\begin{eqnarray}
\cfrac{\langle\delta \hat{\mathrm{X}}^2\rangle} {\langle\delta \hat{\mathrm{X}}_{\text{vac}}^2\rangle}
=
\text{exp} \Bigg(
{-2 L \, \sqrt{\eta P_{\text{in}}}\, \text{tanh} \Big( L\, \sqrt{\eta\, P_{\text{in}}} \Big) } 
\Bigg).
\label{eq:eq1_SI}
\end{eqnarray}

\begingroup
\setlength{\tabcolsep}{3.5pt} %
\renewcommand{\arraystretch}{1.1} %
\begin{table*}[t]
\begin{tabular}{c|c|c}
Loss source & Loss (dB) & Determination\\ \hline\hline
Waveguide propagation & 0.80 & $Q$-factor measurement in a ring resonator on the same chip \\
Lensed multimode fiber collection & 5.40 & Coupling test to a straight waveguide \\
Detector quantum efficiency & 0.80 & Photodetector documentation
\end{tabular}
\caption{
\textbf{Summary of the sources of loss in the detection chain.}
}
\label{table:tab1}
\end{table*}
\endgroup

\section*{Coherent leakage into the Squeezer and LO phase calibration}

In our system, we observe non-perfect filtering of the squeezer subsection pump at the fundamental frequency. This results from the limited extinction ratio of the filter used on the chip to separate FH from SH after the SHG section. This section explains how we quantify this leakage and use it to calibrate the local oscillator phase.

Our model consists of a tunable beam splitter with two inputs (LO and squeezer output) and two outputs (BHD ports). We calculate a result of a phase modulation imposed on the LO arm on the BHD measurement result, depending on the leakage level. The input state of the system is:
\begin{eqnarray}
     \begin{bmatrix}
         A_{in}^{(1)} \\ 
         A_{in}^{(2)}    
     \end{bmatrix}
      =
     \begin{bmatrix}
         \sqrt{\varepsilon}\alpha_{\text{LO}}(t) + \hat{a}   \\ 
         \alpha_{\text{LO}}(t)e^{i\phi_{\text{LO}}}e^{i\phi_{\text{M}}(t)} + \hat{v}
     \end{bmatrix}.
\end{eqnarray} Here, $\varepsilon$ is the power ratio of the coherent leakage to the LO, $\phi_{\text{LO}}$ is the LO phase with respect to the signal in the squeezer path, $\phi_{\text{M}}$ is the RF phase modulation, $\hat{a}$ and $\hat{v}$ represent quantum fluctuations in the squeezer and LO paths, respectively. We solve the beam splitter matrix equation:
\begin{eqnarray}
     \begin{bmatrix}
         A_{out}^{(1)} \\ 
         A_{out}^{(2)}    
     \end{bmatrix}
      = 
\begin{bmatrix}
\text{sin}(\phi_2/2) & \text{cos}(\phi_2/2) \\
\text{cos}(\phi_2/2) & -\text{sin}(\phi_2/2)
\end{bmatrix}
     \begin{bmatrix}
         A_{in}^{(1)} \\ 
         A_{in}^{(2)}    
     \end{bmatrix}
     \label{eq:matrix_eq}
\end{eqnarray} to find the BHD signal generated. This model is equivalent to the signal interferometer described in the main text. For no power leakage, it is biased such that the phase is around $\phi_2 = \pi/2$ but $\varepsilon \neq 0$ results in an imbalance that we have to compensate for.

We solve for the output port amplitudes and use them to find the BHD differential signal defined as \mbox{$P_{\text{BH}} = |A_{out}^{(1)}|^2 - |A_{out}^{(2)}|^2$}. The result is given by:
\begin{widetext}
\begin{eqnarray} \nonumber
P_{\text{BH}} &=& 
\text{cos}(\phi_2) |\alpha_{\text{LO}}|^2 (1-\varepsilon) 
+
2\text{sin}(\phi_2)\sqrt{\varepsilon} |\alpha_{\text{LO}}|^2 \text{cos}(\phi_{\text{LO}}) 
\\ \nonumber
&+& \text{cos}(\phi_2) [\alpha_{\text{LO}} e^{i\phi_{\text{LO}}} \hat{v}^{\dagger} - \sqrt{\varepsilon} \alpha_{\text{LO}} \hat{a}^{\dagger} + {h.c} ] 
\\
&+& \text{sin}(\phi_2) [\alpha_{\text{LO}} e^{i\phi_{\text{LO}}} \hat{a}^{\dagger} + \sqrt{\varepsilon} \alpha_{\text{LO}} \hat{v}^{\dagger} + {h.c} ]
\label{eq:BHD_sig}
\end{eqnarray} 
\end{widetext}
We separate equation \ref{eq:BHD_sig} into DC and AC parts. The former defines the locking condition:
\begin{eqnarray}
\text{cos}(\phi_2)  (1-\varepsilon) 
+
2\text{sin}(\phi_2)\sqrt{\varepsilon} \text{cos}(\phi_{\text{LO}})
= 0 
\label{eq:lock_condition},
\end{eqnarray} which is solvable analytically:
\begin{eqnarray} 
\phi_{2} =
\pm
\text{arccos}
\left[
\pm \cfrac{2\sqrt{\varepsilon} \text{cos}(\phi_{\text{LO}}) }
{\sqrt{4\varepsilon \text{cos}^2(\phi_{\text{LO}}) + \varepsilon^2-2\varepsilon+1  }}
\right]
\label{eq:locked_phase}.
\end{eqnarray}

In addition to locking, we use the DC term to calibrate the LO phase and directly measure the leakage $\varepsilon$. For this purpose, we introduce a periodic modulation to the LO phase shift, such that $\phi_{\text{LO}} = \phi_{\text{LO}} + \phi_{\text{M}}(t) = \phi_{\text{LO}} + \pi \, V_{\text{p-p}}/(2V_{\pi}) \, \text{sin}(\Omega t)$. $\Omega$ is the modulation frequency (here set to 72~MHz), $V_{\pi}$ is the half-wave voltage, and $V_{\text{p-p}}$ is the peak-to-peak modulation voltage. We focus on the DC part of the equation \ref{eq:BHD_sig}, apply a standard Taylor series expansion to the phase modulation term, and write the resulting signal in the frequency domain at the peak frequency of the modulation:
\begin{eqnarray} 
P_{\text{RF}}(\Omega)
=
\cfrac{2\varepsilon R^2 P_{\text{LO}}^2}{Z}
\left(
\cfrac{\pi}{2}
\right)^3
\cfrac{V_{\text{p-p}}}{V_{\pi}}
\text{sin}^2
\left(
\cfrac{V_{\text{DC}}}{V_{\pi}} \pi
\right),
\label{eq:cal_sig}
\end{eqnarray} $V_{\text{DC}}$ is the DC voltage bias at the LO phase shifter. Signal strength depends on the local oscillator power $P_{\text{LO}}$, detection responsivity $R$, the impedance of the detector $Z$, and the leakage $\varepsilon$. We fit equation \ref{eq:cal_sig} in the main text to find $\varepsilon$ and $V_{\pi}$.

\section*{Impact of leakage on the squeezing visibility} 

The AC part of the equation \ref{eq:BHD_sig} probes both the noise of the prepared quantum state and the noise of the LO. It results in the noise measured in the main text:
\begin{eqnarray} \nonumber
P_{\text{BH}}^{\text{(AC)}}
&=& \text{cos}(\phi_2) [\alpha_{\text{LO}} e^{i\phi_{\text{LO}}} \hat{v}^{\dagger} - \sqrt{\varepsilon} \alpha_{\text{LO}} \hat{a}^{\dagger} + {h.c} ]
\\ \nonumber
&+& \text{sin}(\phi_2) [\alpha_{\text{LO}} e^{i\phi_{\text{LO}}} \hat{a}^{\dagger} + \sqrt{\varepsilon} \alpha_{\text{LO}} \hat{v}^{\dagger} + {h.c} ].
\\
\label{eq:BHD_sig_AC}
\end{eqnarray} Note that it reduces to the usual BHD expression for $\varepsilon = 0$ and $\phi_2 = \pi/2$. Our experiment introduces two modifications to the classical BHD picture. One is the leakage $\varepsilon \neq 0$, which results in probing the local oscillator noise $\hat{v}$, in addition to measuring the squeezed state $\hat{a}$. The other modification is varying the splitting ratio of the output beam splitter $\phi_2 \neq \pi/2$. This adds a term proportional to $\text{cos}(\phi_2)$, resulting in changing the measured noise characteristics. As a result, measured squeezing can be reduced. This section explains how the power leakage $\varepsilon$ impacts observable squeezing. We first rewrite the equation \ref{eq:BHD_sig_AC} with the quadrature operators for the squeezing and LO paths \mbox{$\hat{a} = \delta \hat{\mathrm{X}} + i \delta \hat{\mathrm{Y}}$} and \mbox{$\hat{v} = \delta \hat{\mathrm{X}}_{\text{LO}} + i \delta \hat{\mathrm{Y}}_{\text{LO}}$}:
\begin{eqnarray} \nonumber
P_{\text{DIFF}}^{\text{(AC)}} &=& 
2\,\text{sin}(\phi_2)\, |\alpha_{\text{LO}}| \times
\\ \nonumber
&&
\left[
\text{cos}(\phi_{\text{LO}}) \delta\hat{\mathrm{X}} \right.
+
\left.
\text{sin}(\phi_{\text{LO}}) \delta\hat{\mathrm{Y}}
+
2\sqrt{\varepsilon}\delta\hat{\mathrm{X}}_{\text{LO}}.
\right]
\\ \nonumber
&+&
2\, \text{cos}(\phi_2)\, |\alpha_{\text{LO}}| \times
\\ \nonumber
&&
\left[
\text{cos}(\phi_{\text{LO}}) \delta\hat{\mathrm{X}}_{\text{LO}} \right.
+ \left.
\text{sin}(\phi_{\text{LO}}) \delta\hat{\mathrm{Y}}_{\text{LO}}
-
2\,\sqrt{\varepsilon}\delta\hat{\mathrm{X}} \right]
\\
\label{eq:BHD_noRF}
\end{eqnarray} This is an expected result for the BHD applied toward phase-sensitive probing of the quadratures of a squeezed state with an additional term that probes the noise of the LO. We assume the LO noise to be phase-insensitive and shot-noise-limited. The signal variance is proportional to the power measured by the spectrum analyzer and given by equation \ref{eq:BHD_with_leakage}. We use this result as a model for our squeezing measurement in the main text.
\begin{widetext}
\begin{eqnarray} \nonumber
\langle 
\delta P_{\text{DIFF}}^{\text{(AC)}2}
\rangle
&=&
4\, |\alpha_{\text{LO}}|^2
\left[\,
\text{cos}^2(\phi_2)
\left( \,
\text{cos}^2(\phi_{\text{LO}}) \langle 
\delta \hat{\mathrm{X}}_{\text{LO1}}^2
\rangle
+\text{sin}^2(\phi_{\text{LO}}) \langle 
\delta \hat{\mathrm{X}}_{\text{LO2}}^2
\rangle
+
4\,\varepsilon \, \langle 
\delta \hat{\mathrm{X}}_{\text{1}}^2
\rangle
\,\right)
\right.
\\ \nonumber
&+&
\text{sin}^2(\phi_2)
\left(\,
\text{cos}^2(\phi_{\text{LO}}) \langle 
\delta \hat{\mathrm{X}}_{\text{1}}^2
\rangle
+\text{sin}^2(\phi_{\text{LO}}) \langle 
\delta \hat{\mathrm{X}}_{\text{2}}^2
\rangle
+
4\varepsilon\langle 
\delta \hat{\mathrm{X}}_{\text{\text{LO}1}}^2
\rangle
\,\right)
\\
&+& \left.
2\,\text{sin}(2\phi_{2}) \,
\sqrt{\varepsilon} \,
\text{cos}(\phi_{\text{LO}})
\left( \,
\langle 
\delta \hat{\mathrm{X}}_{\text{\text{LO}1}}^2
\rangle -
\langle 
\delta \hat{\mathrm{X}}_{\text{1}}^2
\rangle
\,\right)  \,\right]
\label{eq:BHD_with_leakage}
\end{eqnarray}
\end{widetext}

\section*{Effects of loss in the system}

The total detection chain efficiency is a major challenge in characterizing squeezed light. This is especially consequential when working with complex PICs like the one described in the main text. In our experiment the total detection chain efficiency was around $\zeta \approx \text{20}\%$, this leads to the reduction of squeezing visibility:
\begin{eqnarray}
\left[\cfrac{\langle\delta \hat{\mathrm{X}}^2\rangle} {\langle\delta \hat{\mathrm{X}}_{\text{vac}}^2\rangle}
\right]_{\text{}}
=
\left(
\zeta
\left[\cfrac{\langle\delta \hat{\mathrm{X}}^2\rangle} {\langle\delta \hat{\mathrm{X}}_{\text{vac}}^2\rangle}
\right]_{\text{On-Chip}}
+(1-\zeta)
\right).
\label{eq:squeezing_vs_detection_eff}
\end{eqnarray} We summarize all the sources of loss that impact reported total detection efficiency in table \ref{table:tab1}. In our measurement, the main limitation is a result of the loss at the interface between the chip facet and collection fibers.

\begin{figure}
  \begin{center}
      \includegraphics[width=1\columnwidth]{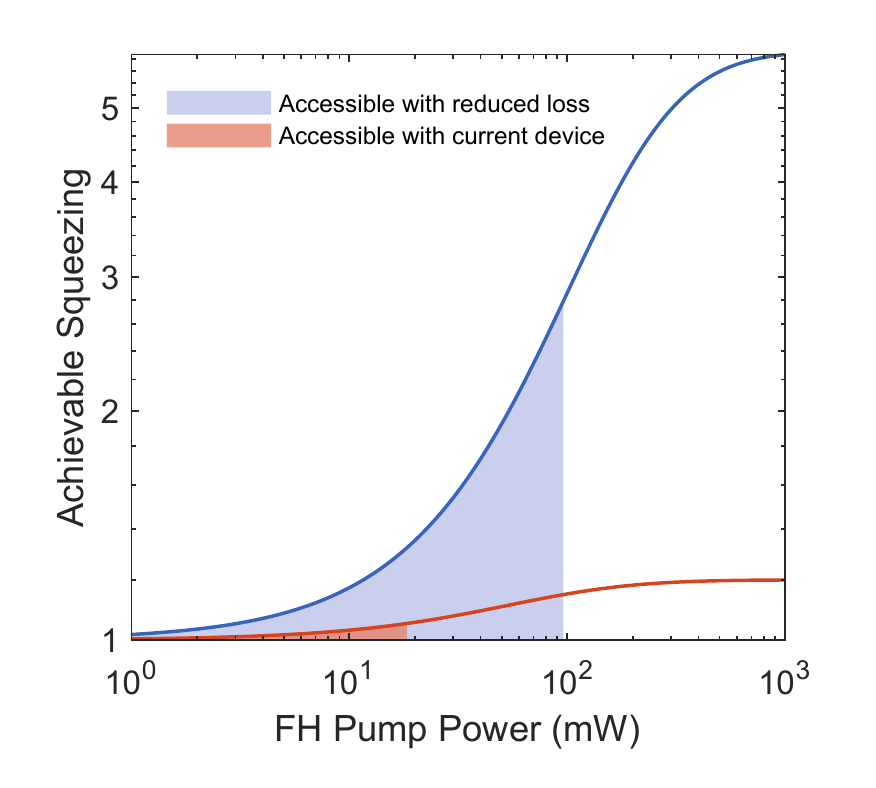}
  \end{center}
 \caption{
 \textbf{Calculated squeezing with reduced loss. }
We calculate the expected squeezing for the proposed device with reduced detection and laser coupling loss. We assume a total laser coupling loss of 0.6 dB and a total detection loss of 0.8 dB. The blue line corresponds to the expected squeezing after reducing optical loss in the system. The orange line corresponds to the squeezing achievable with the current system, with 20$\%$ detection efficiency. The shaded areas are accessible with the current integrated DFB laser for both cases.
}
\label{fig:figSI3}
\end{figure}

Reducing the overall losses in the system is a straightforward way to improve the device's performance. Reducing the fiber-to-chip coupling loss down to 0.6 dB has been established by engineering the waveguide termination.~\cite{He2019, Hu2021, Ying2021} Moreover, the propagation loss can be reduced to 0.2 dB by reducing the waveguide length where the squeezed state can experience loss. The quantum efficiency of the detector used in this study can be improved to 99.5$\%$ using commercially available photodiodes. We use these values to estimate potentially achievable squeezing in the proposed device. In this case, the total detection loss is 0.8 dB, and the total insertion loss of the laser is 0.6 dB. Using these values, we plot the expected achievable squeezing with respect to the laser power in Fig.~\ref{fig:figSI3}. Assuming no squeezer leakage, we compare the calculated values in a system with reduced loss (blue) to the current system with a total detection efficiency of around 20$\%$ (orange). We estimate that the same integrated DFB laser, as we used in the main text, would facilitate access to the blue-shaded region and result in the maximum squeezing factor of 2.8 at the highest achievable on-chip power of 96 mW.


\end{document}